\documentclass[10pt,a4paper,twocolumn]{article}
\usepackage[utf8]{inputenc}
\usepackage[T1]{fontenc}
\usepackage{lmodern}
\usepackage[english]{babel}
\usepackage{microtype}
\usepackage[top=2cm,bottom=2.5cm,left=2cm,right=2cm,columnsep=8mm]{geometry}
\usepackage{amsmath,amssymb}
\usepackage{graphicx}
\usepackage{multirow}
\usepackage[font=small,labelfont=bf]{caption}
\usepackage{url}
\usepackage[colorlinks=true,linkcolor=blue,citecolor=blue,urlcolor=blue]{hyperref}

\newcommand{\keywords}[1]{\par\smallskip\noindent\small\textbf{Keywords:} \textit{#1}\par\normalsize}

\title{Predicting Audio Popularity: A Comparative Study of Metadata and Acoustic Features Across Music and Sound Effects}

\author{
Ikram El Hadi$^{1,*}$\enspace Nihad Elbari$^{1,*}$\enspace Kawtar Tlem\c{c}ani$^{1,*}$\enspace Th\'eo Mariotte$^{2}$\enspace Meysam Shamsi$^{1,2}$\\[0.5em]
\normalsize\textit{$^{1}$ENSIM, Le Mans Universit\'e, France}\\
\normalsize\textit{$^{2}$LIUM, Le Mans Universit\'e, France}\\[0.3em]
\normalsize\texttt{meysam.shamsi@univ-lemans.fr}\\[0.3em]
\small $^{*}$These authors contributed equally to this work.
}
\date{}

\hypersetup{pdftitle={Predicting Audio Popularity: A Comparative Study of Metadata and Acoustic Features Across Music and Sound Effects}}

\begin{document}
\maketitle

\begin{abstract}
This study addresses the prediction of audio popularity on an online sharing platform by leveraging both metadata and acoustic feature representations. Two distinct categories of audio content are considered: music and sound effects. Popularity is formulated as a supervised classification task based on two target variables: the number of downloads and the average user rating.
We conduct a comparative evaluation of multiple machine learning models, including Random Forest, XGBoost, and a Neural Network model. Experiments are performed across different feature sets, including metadata, platform-provided descriptors, and eGeMAPS acoustic features.
Results indicate that metadata consistently provide the most informative representation for predicting popularity. Nevertheless, acoustic features alone achieve performance significantly above chance level, demonstrating that intrinsic properties of the audio signal contribute to popularity prediction.

\end{abstract}

\keywords{audio popularity, machine learning, metadata, acoustic features}

\section{Introduction}

The rapid growth of audio content on online sharing platforms has made popularity prediction a key challenge for recommendation systems and user behavior analysis. Accurately forecasting the potential success of audio items remains a complex task, as it depends on a combination of intrinsic content characteristics and user interaction dynamics. Prior studies \cite{lee2018music,khan2022feature,martin2020multimodal} have demonstrated that both metadata (e.g., tags, uploader identity, and upload time) and acoustic features can partially explain popularity patterns. However, a substantial portion of variability is still driven by external and often unobservable factors, such as social influence and platform-specific dynamics.

In this work, we investigate the prediction of audio popularity using a dataset collected from an online audio platform. Popularity is formulated as a supervised classification problem based on platform-derived user engagement indicators. We consider two distinct categories of audio content—music and sound effects—in order to analyze potential differences across domains characterized by different structural and usage properties.

To this end, we conduct a comparative evaluation of multiple machine learning models, including tree-based methods and neural networks. We examine the contribution of different feature families—metadata and acoustic representations—as well as the impact of class imbalance and its mitigation through oversampling techniques. Model performance is assessed using a cross-validated evaluation protocol to ensure robustness and comparability across experimental settings.

\section{Related Work}

The prediction of music popularity has attracted significant attention in recent years, particularly in the context of streaming and online audio platforms. Popularity is typically quantified using metrics such as the number of streams, downloads, or platform-specific scores \cite{lee2018music, interiano2018musical}. These indicators can be modeled either as continuous variables or discretized to formulate the problem as a classification task.

A substantial body of work has focused on leveraging acoustic features to predict popularity. Lee and Lee \cite{lee2018music} investigate the relationship between automatically extracted audio descriptors, such as tempo, energy, and key, and success indicators derived from commercial platforms. Their approach relies on statistical analysis and supervised learning models to identify the most influential features. Similarly, Khan et al. \cite{khan2022feature} compare different sets of audio descriptors and evaluate their impact on classification performance, showing that feature selection plays a critical role in prediction accuracy. In another line of work, Cimen et al. \cite{cimen2021longitudinal} propose a longitudinal approach that models the evolution of popularity over time by incorporating temporal information, enabling a better understanding of success dynamics.

Other studies adopt a multimodal perspective. Martin et al. \cite{martin2020multimodal} combine acoustic features with metadata as well as textual data (e.g., from social media) to improve predictive performance. These approaches consistently demonstrate that integrating multiple data sources leads to better results than relying on a single modality.

However, most existing work relies on proprietary platforms or datasets that do not provide access to raw audio signals. Relatively few studies exploit open platforms, which offer simultaneous access to audio files, metadata and popularity indicators. Moreover, systematic comparisons between metadata-based and purely acoustic approaches remain limited.

In this context, the present work contributes a comparative analysis of different feature types, explicitly evaluating the predictive power of acoustic features alone, as well as the impact of class imbalance and its mitigation on model performance.

\section{Data collection}

In this study, we focus on two distinct categories of audio content: music and sound effects, denoted as \textbf{Mu} (Music) and \textbf{SE} (SoundEffect), respectively. This categorization is motivated by differences in usage, context, and structural properties between the two types. Music tracks are typically longer and organized around harmonic and rhythmic structures, while sound effects are short, event-driven, and often used as discrete auditory cues. The distinction is also relevant for feature extraction, as different acoustic descriptors may be more informative for one category than the other—for example, tempo and key features are more meaningful for music, whereas temporal envelopes and spectral bursts are more relevant for sound effects.

The data used in this study consist of audio files collected from an online audio-sharing environment FreeSound\footnote{\url{https://freesound.org/docs/api/}}
that provides access to raw audio signals, descriptive metadata, and user interaction indicators. Such environments allow large-scale analysis by combining content-level information with user feedback (e.g., downloads, ratings).

We propose to quantify the popularity using two main variables: the number of downloads and the average user rating. The number of downloads reflects the overall level of user interest, while the average rating provides a qualitative assessment when available.

Both variables exhibit highly skewed distributions, a common characteristic of digital content platforms, where a majority of items receive limited attention while a small fraction concentrates most interactions.

To formulate popularity prediction as a supervised learning problem, these continuous variables are discretized into categorical classes.
The number of downloads is divided into three ordinal classes using quantiles, corresponding to low, medium, and high popularity levels. This approach ensures balanced class distributions while preserving the natural ordering of the variable.
The average rating is transformed into four classes. Files without any ratings are grouped into a dedicated class labeled 0 (\textit{MissingInfo}). Rated files are further divided into three ordinal classes representing low, medium, and high appreciation levels. This formulation explicitly distinguishes between the absence of ratings and actual user evaluations.

The resulting task consists of predicting, for a given audio file, two target variables: a popularity class based on the number of downloads (three levels) and a rating class based on the average score (four levels), while accounting for the type of audio content.

Audio duration distributions differ markedly between music and sound effects. Music tracks (10–60 seconds) are positively skewed, with most concentrated around shorter durations (10–20 seconds) and a long tail toward longer lengths.
Sound effects are much shorter (0.5–3 seconds) and exhibit a compact distribution centered on low durations, reflecting their transient nature. This clear structural difference justifies analyzing the two categories separately.

\subsection{Number of downloads}
The \textit{num-downloads} variable shows strong skewness in both categories. Most files accumulate a small number of downloads, while a limited subset reaches very high values. This pattern is characteristic of long-tail distributions commonly observed in digital content ecosystems and is particularly pronounced for \textit{SoundEffect} items.

\subsection{Average rating}
The \textit{avg-rating} variable exhibits a strong concentration around high values (close to 5) for both \textit{Music} and \textit{SoundEffect}. Low ratings are relatively rare, while most rated items receive high scores. This suggests the presence of a positive rating bias, frequently observed in collaborative environments. Additionally, a substantial proportion of files has no ratings, further contributing to class imbalance after discretization.

\section{Feature Extraction}

Each audio file is represented using two complementary types of information: (i) metadata associated with the content, and (ii) acoustic features extracted from the raw audio signal.

Acoustic features are further grouped into three complementary sets:
(1) descriptors provided by the FreeSound platform,
(2) descriptors extracted using the openSMILE toolkit (eGeMAPS), and
(3) representations obtained from a self-supervised learning (SSL) model.

\subsection{Metadata}
Metadata correspond to descriptive information associated with each audio file. They include general attributes such as duration, sampling rate, as well as textual or categorical information (e.g., categories and tags). Metadata provide contextual information about the nature of the content and its potential use, and can be particularly informative for predicting popularity and user preferences.

\subsection{Acoustic Features}

\subsubsection{Platform-provided descriptors Acu-FS}
The FreeSound platform computes a set of acoustic descriptors directly from the audio signal. These features capture low-level properties covering spectral, temporal, and harmonic aspects of the sound, providing a compact representation of timbre and structural characteristics.

\subsubsection{eGeMAPS descriptors}
Additional acoustic features are extracted using the eGeMAPS toolkit, originally developed for speech emotion recognition \cite{eyben2010opensmile}.
This feature set offers a compact and perceptually meaningful set of descriptors.

These features include measures related to fundamental frequency (pitch), energy, and spectral properties. Features are aggregated into \textit{functionals}, i.e., statistical summaries computed over the entire audio signal, producing a fixed-length vector for each file.

\subsubsection{Self-Supervised Learning Representations}
High-level audio representations are extracted using the MERT model (\textit{Music Encoder Representations from Transformers}) \cite{li2023mert}, a foundation model based on the Transformer architecture and pre-trained on large-scale musical corpora.

Unlike traditional acoustic descriptors, MERT learns abstract, high-level representations directly from raw audio in a self-supervised manner. These embeddings capture rhythmic, harmonic, and timbral structures that go beyond low-level signal properties. This approach is particularly suitable for music content, although it can also provide informative features for other types of audio.
Embeddings derived from this model are treated as a distinct feature set and evaluated separately from the other descriptors to compare their predictive power for popularity.

\section{Methodology}

This section presents the experimental protocol for predicting the popularity of audio content using the different feature representations described previously. It details the preprocessing applied to each feature type, the models used, and the evaluation strategies.

Experiments are conducted separately for music and sound effects to analyze the influence of content type on model performance. For each feature set, preprocessing steps are applied consistently, followed by model training and evaluation under a unified protocol, enabling a fair comparison across approaches.

\subsection{Data Preprocessing}

\subsubsection{Metadata}

Metadata undergo structured preprocessing before being used for learning. Non-informative columns (e.g., file paths or preview references) are removed to reduce dimensionality and noise.

Metadata is preprocessed into a unified numerical format. Skewed numerical variables (e.g., ratings, comments, duration) are log-transformed, while others are standardized. Temporal data is converted into an age feature to capture recency effects. Categorical variables are one-hot encoded, with rare values grouped, and boolean fields binarized. Contributor activity is represented by frequency of contributions. Text features are vectorized: tags as multi-label indicators, filenames via hashing plus length, and descriptions using averaged GloVe \cite{pennington2014glove} embeddings. The final metadata representation consists of 22 numerical features per audio sample.

These transformations produce a homogeneous numerical representation of metadata suitable for classification models. In total, the metadata representation includes 22 numerical variables describing each audio sample.

\subsubsection{Acoustic Features}

\textbf{Acu-FS}:
Low-level acoustic descriptors provided by the platform are retained as a first representation of the audio signal (\textit{Acu-FS}). Only numerical features are used, while unique identifiers and target variables are excluded. Missing values are imputed with zero, representing unavailable information. These descriptors are used to evaluate the predictive power of intrinsic signal properties. The resulting representation consists of 47 numerical variables per audio sample.
These descriptors are mainly calculated using the Essentia\footnote{\url{https://essentia.upf.edu/}} toolkit.

\textbf{eGeMAPS}:
The eGeMAPSv02 features are extracted at the \textit{functional} level \cite{eyben2016egemaps}, producing an 88-dimensional vector per file. Features include fundamental frequency (F0), cepstral coefficients (MFCC 1–4), formant frequencies (F1–F3), spectral parameters (slope, spectral flux, Hammarberg index), rhythmic indicators (voiced/unvoiced), loudness, and voice quality measures (jitter, shimmer, harmonic-to-noise ratio).
Aggregated statistics (\textit{functionals} such as mean, standard deviation, percentiles) provide fixed-length vectors independent of audio duration. No additional preprocessing is applied before classification.

\textbf{MERT}~\cite{li2023mert}:
Each audio file is first preprocessed to ensure consistency across inputs. Audio signals are converted to mono and resampled to 24000~Hz. To standardize input duration, signals are truncated to a maximum predefined length.
The preprocessed waveform is then passed through the MERT model (\textit{MERT-v1-330M}). The resulting latent representation is obtained from the last hidden layer of the network. A temporal average pooling is applied over the sequence dimension to produce a fixed-length embedding vector for each audio file. Each audio sample is thus represented by a dense numerical vector of dimension 1024.
These embeddings are used directly as input features for downstream classification -- popularity prediction -- without additional preprocessing.

\subsection{Models and Experimental Protocol}

Popularity prediction is performed using three supervised learning models: Random Forest,
XGBoost,
and a Multi-Layer Perceptron (MLP). These models are well-suited for high-dimensional, heterogeneous, and potentially nonlinear data, and are robust to varying feature distributions.

The Random Forest and XGBoost models are tree-based ensemble methods that can effectively capture nonlinear relationships and interactions between features while requiring minimal preprocessing.

In addition, a feedforward neural network is considered to evaluate the performance of deep learning approaches on the task. The MLP consists of three hidden layers with sizes (128, 64, 32), using ReLU nonlinear activation functions. To ensure proper scaling of input features, the model is implemented within a pipeline including mean and variance standardization ($Z$-norm). The network is trained using a maximum of 300 epochs, with early stopping enabled to prevent overfitting.

Experiments are conducted separately for music and sound effects to examine content-specific performance differences.
For each feature set, data are split into training and test subsets according to the protocol of Table~\ref{tab:splits}, ensuring fair comparison across representations.
Due to the strong class imbalance in popularity variables, SMOTE~\cite{chawla2002smote} is applied only to training sets to oversample minority classes. This improves the ability of models to predict underrepresented classes.

\begin{table}[t]
\centering
\caption{Number of files in Train and Test splits. Ratios represent the proportion of data kept in a given split from the original dataset. Both sets are disjoint.}
\begin{tabular}{lcc}
\hline
& \textbf{SE} & \textbf{Mu} \\
\hline
Train (80\%) & 13,963 & 14,340 \\
Test (20\%) & 3,491 & 3,586 \\
\hline
\end{tabular}
\label{tab:splits}
\end{table}

Model performance is evaluated using F1-score, computed separately for each target variable (number of downloads and average rating). This protocol allows comparison of model type, feature set and audio domain -- music or sound event -- for audio content popularity prediction.

\section{Results}

Table~\ref{tab:results_all_features} summarizes the classification performance (F1-score) obtained across feature sets, models, and targets for each audio domain.

\begin{table*}
\caption{Classification performance F1-score ($\pm$95\% CI). \textbf{Bold} values indicate the best model and \underline{underlined} values the second best.}
\label{tab:music_combined}
\resizebox{\textwidth}{!}{
\begin{tabular}{l|l|lll|lll}
\hline
 & Target & \multicolumn{3}{c}{Avg-Rat.} & \multicolumn{3}{|c}{\#Down.} \\
\hline
Dataset &  Model & MLP & RandomForest & XGBoost & MLP & RandomForest & XGBoost \\
\hline
\multirow{4}*{\textbf{Mu}} & Acu-FS & \textbf{0.441} $\pm$ 0.005 & \textbf{0.562} $\pm$ 0.010 & \textbf{0.640} $\pm$ 0.014 & \underline{0.586} $\pm$ 0.019 & \underline{0.644} $\pm$ 0.012 & \underline{0.657} $\pm$ 0.010 \\
& eGeMAPS & 0.407 $\pm$ 0.008 & 0.410 $\pm$ 0.009 & 0.414 $\pm$ 0.010 & 0.568 $\pm$ 0.008 & 0.601 $\pm$ 0.008 & 0.594 $\pm$ 0.006 \\
& MERT & 0.303 $\pm$ 0.011 & 0.313 $\pm$ 0.011 & 0.315 $\pm$ 0.013 & 0.461 $\pm$ 0.006 & 0.489 $\pm$ 0.012 & 0.479 $\pm$ 0.008 \\
& Metadata & \underline{0.411} $\pm$ 0.015 & \underline{0.472} $\pm$ 0.015 & \underline{0.510} $\pm$ 0.017 & \textbf{0.623} $\pm$ 0.006 & \textbf{0.732} $\pm$ 0.004 & \textbf{0.735} $\pm$ 0.005 \\
\hline
\multirow{4}*{\textbf{SE}} & Acu-FS & \underline{0.238} $\pm$ 0.012 & \textbf{0.487} $\pm$ 0.010 & \textbf{0.572} $\pm$ 0.019 & \underline{0.550} $\pm$ 0.016 & \underline{0.706} $\pm$ 0.003 & \underline{0.720} $\pm$ 0.005 \\
& eGeMAPS & 0.223 $\pm$ 0.000 & 0.223 $\pm$ 0.000 & 0.223 $\pm$ 0.000 & 0.203 $\pm$ 0.000 & 0.203 $\pm$ 0.000 & 0.203 $\pm$ 0.000 \\
& MERT & 0.223 $\pm$ 0.000 & 0.223 $\pm$ 0.000 & 0.223 $\pm$ 0.000 & 0.203 $\pm$ 0.000 & 0.203 $\pm$ 0.000 & 0.203 $\pm$ 0.000 \\
& Metadata & \textbf{0.292} $\pm$ 0.048 & \underline{0.314} $\pm$ 0.013 & \underline{0.411} $\pm$ 0.011 & \textbf{0.722} $\pm$ 0.018 & \textbf{0.822} $\pm$ 0.004 & \textbf{0.846} $\pm$ 0.004 \\
\hline
\end{tabular}
}
\label{tab:results_all_features}
\end{table*}

Across both datasets, metadata emerge as the most reliable feature set, particularly for predicting the number of downloads. Acoustic features (Acu-FS) provide complementary information and are especially effective for average rating prediction in music and moderately effective in sound effects. In contrast, eGeMAPS and MERT features show limited or inconsistent performance, particularly for sound effects, where they fail to capture meaningful distinctions in popularity.

\subsection{Music dataset (Mu)}

\paragraph{Average rating.}
For predicting average rating on music, platform-provided acoustic features (Acu-FS) yield the strongest performance, reaching an F1-score of $0.64$. Metadata follow with a slightly lower score of $0.51$, while eGeMAPS ($0.41$) and MERT ($0.32$) lag behind. This suggests that perceptually grounded acoustic descriptors are more informative than high-level metadata for this target.

\paragraph{Number of downloads.}
For download prediction, metadata clearly outperform all other feature sets, achieving an F1-score of $0.74$. Acoustic features (Acu-FS) provide competitive but lower performance ($0.66$), while eGeMAPS ($0.59$) and MERT ($0.49$) remain less effective. This indicates that popularity in terms of downloads is more strongly associated with contextual and descriptive information than with audio content alone.

\subsection{Sound effects dataset (SE)}

Predicting popularity for sound effects using external feature extractors such as eGeMAPS and MERT proves particularly challenging. These approaches rely on automatic extraction of generic acoustic representations that are designed to be task-agnostic (at least in our case). However, sound effects are typically short and event-based signals, which limits the amount of information available for deriving stable and discriminative features. As a result, these representations may fail to capture meaningful characteristics relevant to popularity, especially when compared to metadata or platform-specific descriptors.

Performance for average rating prediction on sound effects is generally lower. Metadata again provide the best results, reaching an F1-score of $0.41$. Acoustic features (Acu-FS) achieve moderate performance ($0.57$), outperforming metadata in this specific case, while eGeMAPS and MERT collapse to baseline levels ($0.22$), indicating a lack of discriminative power for this target.

For download prediction, metadata substantially outperform all other feature sets, achieving the highest F1-score of $0.85$. Acoustic features (Acu-FS) follow with $0.72$, showing solid predictive capability. In contrast, eGeMAPS and MERT again remain at baseline ($0.20$), confirming their limited usefulness for this dataset.

\section{Discussion}

The results indicate a clear distinction in the predictive value of different feature types for popularity measures. Metadata emerge as the most informative source for predicting the number of downloads, achieving the highest F1-scores across both music and sound effects datasets. This underscores that contextual and descriptive information—such as file attributes, tags, or platform-related metadata—plays a central role in driving user adoption.

In contrast, Acu-FS are particularly useful for predicting average rating. Their perceptually grounded descriptors capture intrinsic audio properties that align with listener preferences, making them more effective than metadata for estimating perceived quality. While Acu-FS features are generally less informative than metadata for download-based popularity, they provide complementary insights that are especially relevant for rating prediction.

eGeMAPS and self-supervised embeddings (MERT) show more modest and inconsistent performance across both targets. Despite their richness, these features alone are insufficient for reliable popularity prediction, suggesting that fine-grained acoustic characteristics are only part of the factors influencing user behavior. Overall, the findings highlight a complementary pattern: contextual metadata dominate in predicting user adoption (downloads), whereas acoustic descriptors are more informative for predicting listener evaluation (ratings).

\subsection{Impact of Content Type}

Clear differences emerge between music (Mu) and sound effects (SE). Sound effects show higher and more stable predictive performance, particularly for downloads, where metadata achieve the best results ($0.85$). Acoustic features are more effective for predicting ratings ($0.57$), suggesting that intrinsic signal quality influences perceived quality more than popularity. This consistency likely reflects the functional and concise nature of sound effects.

Music is more challenging to predict. While metadata remain strongest for downloads ($0.74$), acoustic features outperform them for rating prediction ($0.64$ vs. $0.51$), indicating that listener evaluations rely more on perceptual audio characteristics. One explanation is that longer music tracks lose temporal structure when reduced to aggregated acoustic statistics, limiting their effectiveness. Overall, metadata dominate download prediction—likely due to their role in search and discoverability—while acoustic features better capture perceived quality. We hypothesize that users select content primarily based on metadata, whereas ratings reflect the intrinsic acoustic experience.

\subsection{Model Interpretation with SHAP}

To identify the most influential variables, SHAP (SHapley Additive exPlanations)~\cite{lundberg2017shap} was applied to models trained on the sound effects category, focusing on metadata and eGeMAPS descriptors. For each trained XGBoost classifier, we computed SHAP values using the \textit{TreeExplainer}~\cite{janzing2020feature}, which quantifies the contribution of each feature to the model’s predictions on the test set.

For datasets with multiple classes, SHAP values were averaged across classes to obtain a single importance score per feature. In addition, for grouped features such as metadata or acoustic descriptors, individual SHAP values were aggregated according to their original feature grouping, allowing us to interpret the contribution of higher-level feature categories rather than only single variables.

\subsubsection{Metadata}
The SHAP analysis reveals a consistent pattern across both datasets (music and sound effects) and both prediction targets (number of downloads and average rating) for Metadata features. In all configurations, a similar subset of metadata features emerges as the most influential. In particular, \textit{Category} and \textit{SubCategory} show the highest contribution to the predictions, highlighting the importance of semantic classification and content grouping in explaining popularity.

In addition, technical attributes such as \textit{duration} (for music) or \textit{filesize} (for sound effects), along with \textit{license} and \textit{bit depth}, consistently rank among the most important features. These variables capture both usage constraints (e.g., licensing conditions) and signal quality characteristics, which appear to strongly influence user behavior.

\subsubsection{Acu-FS for SE}

The SHAP analysis of Acu-FS highlights a consistent set of perceptually meaningful descriptors that contribute most to the prediction of both average rating and number of downloads. Across both targets, \textit{spectral flatness}, \textit{brightness}, and \textit{temporal spread} emerge as the most influential features for \textbf{SE}.

Spectral flatness, which characterizes the noisiness versus tonality of a signal, appears to play a key role in distinguishing sound textures that may affect both perceived quality and usability.
Similarly, brightness, related to the distribution of energy in higher frequency bands, contributes significantly to the perception of clarity and sharpness, which are important attributes for both evaluation and selection of audio content. Temporal spread, capturing the temporal distribution of energy, provides information about the structure and dynamics of the sound, which may influence how well a sound effect fits a given application context.

\subsubsection{Music dataset (Mu)}

Across Acu-FS features, a consistent pattern emerges: the \textit{warmth} descriptor appears among the top five features for both average rating and number of downloads. Although the notion of warmth is perceptual and not strictly defined in physical terms, it is generally associated with a concentration of energy in low-frequency regions and a smooth spectral profile. Its prominence suggests that such timbral characteristics play an important role in both perceived quality and user preference in music content.

For eGeMAPS features, the analysis also reveals recurrent important descriptors across both targets. In particular, \textit{Mean Unvoiced Segment Length} appears among the top five features. This feature captures temporal properties related to the duration of unvoiced segments, reflecting aspects of articulation and signal structure that may influence both listener perception and engagement.

\section{Conclusion}

In this study, we examined the prediction of audio content popularity across music and sound effects using complementary feature families, including metadata and multiple acoustic representations. The results reveal a clear distinction between prediction targets: metadata are the most informative features for predicting the number of downloads, while platform-provided acoustic features (Acu-FS) are more effective for predicting average rating. This highlights that user adoption is primarily driven by contextual and descriptive information, whereas perceived quality is more closely related to intrinsic audio characteristics.

Furthermore, the analysis shows that generic, automatically extracted representations such as eGeMAPS and MERT provide limited predictive power, particularly for short sound effects, where the lack of sufficient signal duration constrains the extraction of meaningful task-agnostic features.

The SHAP-based interpretation provides additional insight into the models by identifying a small subset of dominant features. Metadata such as \textit{Category}, \textit{SubCategory}, and technical attributes (e.g., \textit{duration}/\textit{filesize}, \textit{license}, \textit{bit depth}) play a central role in prediction, while acoustic descriptors such as spectral flatness, brightness, temporal spread, and perceptual attributes like warmth contribute to capturing signal-related aspects of popularity.

\section*{Generative AI Use Disclosure}
Generative AI was used to edit and polish drafts made by the authors. Any generated content was reviewed and edited by authors who maintain full responsibility for the final content.

\bibliographystyle{ieeetr}
\bibliography{references}

\end{document}